\documentclass{basi}

\usepackage{txfonts}
\usepackage{graphicx}
\newcommand{\CTT}{C^{\rm TT}_\ell}
\newcommand{\CTE}{C^{\rm TE}_\ell}
\newcommand{\CEE}{C^{\rm EE}_\ell}
\newcommand{\CBB}{C^{\rm BB}_\ell}
\newcommand{\CTB}{C^{\rm TB}_\ell}
\newcommand{\CEB}{C^{\rm EB}_\ell}

\def\lsim{\mathrel{\rlap{\lower4pt\hbox{\hskip1pt$\sim$}}
    \raise1pt\hbox{$<$}}}                % less than or approx. symbol
\def\gsim{\mathrel{\rlap{\lower4pt\hbox{\hskip1pt$\sim$}}
    \raise1pt\hbox{$>$}}}                % greater than or approx. symbol

\def\be{\begin{equation}}
\def\ee{\end{equation}}
\def\bea{\begin{eqnarray}}
\def\eea{\end{eqnarray}}

\begin{document}

%% \markboth{Tarun Souradeep}
%% {Cosmology with CMB}

\title[Cosmology with CMB ]{Beyond the Standard cosmological model with CMB}
\author[T. Souradeep]%
       {Tarun Souradeep\thanks{e-mail:tarun@iucaa.ernet.in}\\
        IUCAA, Post Bag 4 , Ganeshkhind, Pune, India}

\pubyear{2011}
%\volume{29}
\pagerange{\pageref{firstpage}--\pageref{lastpage}}
%\setcounter{page}{17}
%\date{Received 2011 April 14; accepted 2011 April 15}

\maketitle

\label{firstpage}

\begin{abstract}
 Measurements of CMB anisotropy and, more recently, polarization have
played a very important role in cosmology. Besides precise
determination of various parameters of the `standard' cosmological
model, observations have also established some important basic tenets
that underlie models of cosmology and structure formation in the
universe -- `acausally' correlated, adiabatic, primordial
perturbations in a flat, statistically isotropic universe.  These are
consistent with the expectation of the paradigm of inflation and the
generic prediction of the simplest realization of inflationary
scenario in the early universe. Further, gravitational instability is
the established mechanism for structure formation from these initial
perturbations. Primordial perturbations observed as the CMB anisotropy
and polarization is the most compelling evidence for new, possibly
fundamental, physics in the early universe. The community is now
looking beyond the parameter estimation of the `standard' model, for
subtle, characteristic signatures of early universe physics.
\end{abstract}


%------------------------------------------------------------------------------%

%% \section{Introduction}	
%%  \label{intro}

The `standard' model of cosmology must not only explain the dynamics
of the homogeneous background universe, but also satisfactorily
describe the perturbed universe -- the generation, evolution and
finally, the formation of large scale structures in the universe.

The transition to precision cosmology has been spearheaded by
measurements of CMB anisotropy and, more recently, polarization.  Our
understanding of cosmology and structure formation necessarily depends
on the rather inaccessible physics of the early universe that provides
the stage for scenarios of inflation (or, related alternatives).  The
CMB anisotropy and polarization contains information about the
hypothesized nature of random primordial/initial metric perturbations
-- (Gaussian) statistics, (nearly scale invariant) power spectrum,
(largely) adiabatic vs.  iso-curvature and (largely) scalar vs. tensor
component.  The `default' settings in brackets are motivated by
inflation.

%% \section{CMB observations and cosmological parameters}
%% \label{cmb}
The angular power spectrum of the Cosmic Microwave Background
temperature fluctuations ($\CTT$)have become invaluable observables
for constraining cosmological models. The position and amplitude of
the peaks and dips of the $\CTT$ are sensitive to important
cosmological parameters, such as, the relative density of matter,
$\Omega_m$; cosmological constant, $\Omega_\Lambda$; baryon content,
$\Omega_B$; Hubble constant, $H_0$ and deviation from flatness
(curvature `density'), $\Omega_K$.
\begin{figure}
\begin{center}
\includegraphics[scale= 0.45]{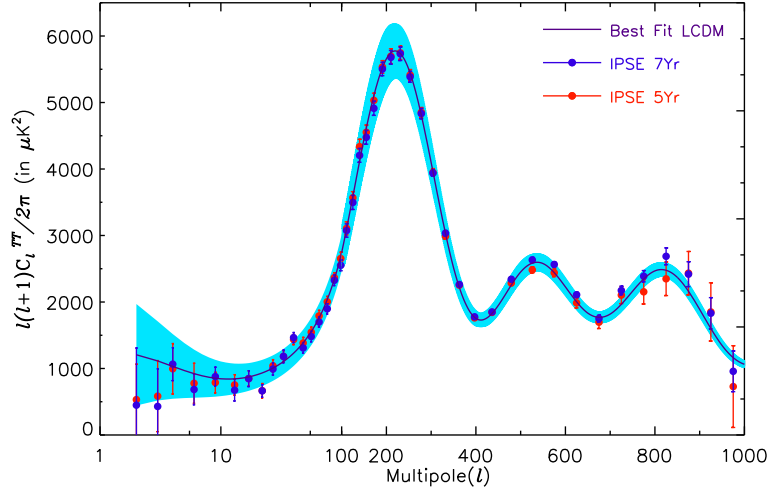}
\caption{ The angular power spectrum estimated from the
multi-frequency five and seven year WMAP data. The result from IPSE a
self-contained model free approach to foreground removal
~\protect{\cite{samal10sah06}} matches that obtained by the WMAP
team. The solid curve showing prediction of the best fit power-law,
flat, $\Lambda$CDM model threads the data points
closely.[Fig. courtesy: Tuhin Ghosh] }
\label{WMAPCL}
\end{center}
\end{figure}

  The angular spectrum, $\CTT$, has been measured with high precision
on up to angular scales ($\ell \sim 1000$) by the WMAP space
mission~\cite{lar_wmap10}, while $\CTT$ at even smaller angular scales
have been measured (at somewhat coarse multipole space resolution) by
ground and balloon-based CMB experiments such as ACBAR, QuaD and
ACT~\cite{quad09,acbar09act10}.  These data are largely consistent
with a $\Lambda$CDM model where the Universe is spatially flat and is
composed of radiation, baryons, neutrinos and, the exotic, cold dark
matter and, at present, dominated by the cosmological constant.
Figure~\ref{WMAPCL} shows the angular power spectrum of CMB
temperature fluctuations obtained from the 5 \& 7-year WMAP
data~\cite{samal10sah06}. Most recent estimates of the cosmological
parameters are available and best obtained from recent literature,
eg. Ref.\cite{lar_wmap10} and, hence, is not given in the article.

One of the firm predictions of this working `standard' cosmological
model is a linear polarization pattern ($Q$ and $U$ Stokes parameters)
imprinted on the CMB at the last scattering surface. The polarization
pattern on the sky can be decomposed in the two kinds, $E$--(gradient)
mode and $B$-- (curl) mode.  For Gaussian CMB sky four power spectra
that characterize the CMB signal~: $\CTT, \CTE, \CEE, \CBB$.  The
expected absence of parity violating physics rule out, $\CTB$ \&
$\CEB$ in usual considerations, however, these are potential probes of
exotic parity violating phenomena.  The CMB polarization spectra
complement $\CTT$ by isolating the effects at the last scattering
surface from that along the line of sight.  Also $\CEE$ provide an
important test on the adiabatic nature of primordial scalar
fluctuations. A clear evidence of adiabatic initial conditions for
primordial density fluctuations is that the compression and
rarefaction peaks in the temperature anisotropy spectrum are `out of
phase' with the gradient (velocity) driven peaks in the $\CEE$.

The first detection of $\CEE$ was achieved by the Degree Angular Scale
Interferometer (DASI) on ($\ell\sim 200-440$) in late
2002~\cite{kov_dasi02}.  First full sky E-mode maps are from
WMAP~\cite{pag_wmap06,kog_wmap03}. The best measurements of $\CTE$ and
$\CEE$ and upper limits on $\CBB$ come from QUaD and
BICEP~\cite{quad09,bicep10}. $\CBB$ is clean probe of early universe
scenarios and required sensitivities at tens of $nK$ level pose stiff
challenges for ongoing and future experiments.

Besides precise determination of various parameters of the `standard'
cosmological model, observations have also begun to establish (or
observationally query) some of the important basic tenets of cosmology
and structure formation in the universe.  The recent cosmological
observations have queried and, in some cases, established fundamental
tenets of cosmology and structure~\footnote{Due to the page limit,
these sections are not expanded upon in this article, for fuller
discussion, see~\cite{me_gr19}. Citations to work carried out in IUCAA
and other important references have, however, been provided.}.:

\begin{itemize}

\item{} {\bf Statistical Isotropy (SI) of the universe:} The {\em
Cosmological Principle} that led to the idealized FRW universe found
its strongest support in the discovery of the (nearly) isotropic,
Planckian, Cosmic Microwave Background.  The exquisite measurement of
the temperature fluctuations in the CMB provide an excellent test bed
for establishing the statistical isotropy (SI) in the universe. The observed CMB sky
is a single realization of the underlying correlation, hence detection
of SI violation, or correlation patterns, pose a great observational
challenge. The Bipolar harmonic representation of CMB sky is emerging
as method of choice for quantifying SI violations~\cite{SI}.

\item{} {\bf Gravitational instability mechanism for structure formation:}
Cosmological perturbations excite acoustic waves in the relativistic
plasma of the early universe. For baryonic density comparable to that
expected from Big Bang nucleosynthesis, acoustic oscillations in the
baryon-photon plasma will also be observably imprinted onto the
late-time power spectrum of the non-relativistic matter. This has been
established (coupled to adiabaticity from CMB polarization results)
through measurements of the subtle Baryon Acoustic Oscillations (BAO) in large
galaxy surveys~\cite{GI}.

\item{} {\bf Origin of primordial perturbations from Inflation:}
What has been truly remarkable is the extent to
which recent cosmological observations have been consistent with and,
in certain cases, even vindicated the simplest set of assumptions for
the initial conditions for the (perturbed) universe.While the simplest
generic inflationary models predict that the spectral index varies slowly with
scale, specific physics in a inflationary model can predict strong scale dependent
fluctuations. Search for subtle features in primordial power spectrum
are being hunted as signatures of new physics~\cite{PPS}.
\end{itemize}

%\section{Conclusions}

The past few years has seen the emergence of a `concordant'
cosmological model that is consistent both with observational
constraints from the background evolution of the universe, as well as, that
from the formation of large scale structures~\cite{me_jpo}.  The
community is now looking beyond the estimation of parameters of a
working `standard' model of cosmology. There is increasing effort
towards establishing the basic principles and assumptions.  The
upcoming results from the Planck space mission will radically improve
the CMB polarization measurements. There are already proposals for the
next generation dedicated satellite mission in 2020+ for CMB
polarization measurements at best achievable sensitivity. The next
decade would see increasing efforts to observationally test
fundamental tenets of the cosmological model and also search for subtle
deviations from the same using the CMB anisotropy and polarization
measurements and related LSS observations, such as, galaxy surveys and
gravitational lensing.

\section*{Acknowledgments}

I would like to thank the organizers for arranging a fruitful
meeting. It is a pleasure to thank and acknowledge the contributions
of students and collaborators who have been involved with the
cosmological quests in the CMB sky at IUCAA.

\end{document}